# Epitaxial growth of highly strained antimonene on Ag (111)

Ya-Hui Mao,[1,†] Li-Fu Zhang,[2,†] Hui-Li Wang,[2] Huan Shan,[1] Xiao-Fang Zhai,[1] Zhen-Peng Hu,[2,*] Ai-Di Zhao[1]* and Bing Wang[1]

[1]*Hefei National Laboratory for Physical Sciences at the Microscale and Synergetic Innovation Center of Quantum Information & Quantum Physics, University of Science and Technology of China, Hefei, Anhui 230026, China*

[2]*School of Physics, Nankai University, Tianjin 300071, China*

[†]These authors contributed equally to this work.

The synthesis of antimonene, which is a promising group-V 2D material for both fundamental studies and technological applications, remains highly challenging. Thus far, it has been synthesized only by exfoliation or growth on a few substrates. In this study, we show that thin layers of antimonene can be grown on Ag (111) by molecular beam epitaxy. High-resolution scanning tunneling microscopy combined with theoretical calculations revealed that the submonolayer Sb deposited on a Ag (111) surface forms a layer of $AgSb_2$ surface alloy upon annealing. Further deposition of Sb on the $AgSb_2$ surface alloy causes an epitaxial layer of Sb to form, which is identified as antimonene with a buckled honeycomb structure. More interestingly, the lattice constant of the epitaxial antimonene (5 Å) is much larger than that of freestanding antimonene, indicating a high tensile strain of more than 20%. This kind of large strain is expected to make the antimonene a highly promising candidate for room-temperature quantum spin Hall material.

## 1 Introduction

Following the successful isolation of graphene in 2004 [1], 2D materials of Xenes (X = Si, Ge, Sn, P, As, Sb, Bi) with similar honeycomb structures have attracted considerable attention from both physics and chemistry communities. However, the zero-bandgap nature of the electronic structures of group-IV Xenes including graphene, silicene, and germanene considerably limits their potential applications in electronics and catalysis. Monoelemental group-V Xenes including phosphorene [2], arsenene [3], antimonene [3], and bismuthene [4] have a buckled honeycomb structure shared with group-IV Xenes such as silicene and germanene [5]. Theory predicted large bandgap openings in these group-V Xenes because of their highly stretchable buckled structure [3]. For example, a bandgap of 2.28 eV is predicted for monolayer antimonene [3]. This kind of large bandgap makes antimonene very promising for potential applications in high-performance electronics and photoelectric devices [6, 7]. More importantly, antimonene under large tensile strain is particularly interesting because it is proposed to be a great candidate for high-temperature quantum spin Hall (QSH) material because of its strong spin-orbital coupling (SOC) effect [8], which is very similar to that of recently discovered bismuthene grown on SiC [9]. Recently, few-layer antimonene has been synthesized by exfoliation [10, 11] and epitaxial growth on mica [12], $PdTe_2$ [13], and germanium surfaces [14]. However, antimonene layers grown on these substrates bear very small tensile strain because these substrates are either van der Waals layered materials or surfaces with a highly matched lattice constant. Experimental evidence of stable existence and stability of highly strained antimonene is scarce and demands immediate verification.

In this study, we present a combined experimental and theoretical investigation on the synthesis and characterization of few layers of antimonene on a Ag (111) surface using scanning tunneling microscopy and first principles calculations. Submonolayer Sb is deposited on the Ag (111) surface and forms a $(\sqrt{3} \times \sqrt{3})R30°$ reconstructed structure upon annealing, which is identified to be a single layer of buckled $AgSb_2$ surface alloy by theoretical calculations. Further deposition of Sb on the $AgSb_2$ surface alloy leads to the formation of epitaxial layers of Sb with a similar $(\sqrt{3} \times \sqrt{3})R30°$ reconstruction under annealing, which is demonstrated to be antimonene layers with a buckled honeycomb structure and an extra-large lattice constant of 5 Å. This kind of large strain in antimonene has not been reported for antimonene grown on other substrates. We attribute the large strain in the antimonene to be stabilized by the buckled $AgSb_2$ surface alloy underneath.

## 2 Experiments and calculations

Experiments were conducted with a low-temperature STM (Scienta Omicron GmbH) with a base pressure of 5 x $10^{-11}$ mbar. The Ag (111) crystal (MaTeck Supplies) was cleaned by cycles of $Ar^+$ sputtering for 15 min at 3.6 x $10^{-6}$ mbar and then annealed at 780 K for 40 min in a preparation chamber (base pressure 1.5 x $10^{-10}$ mbar). Sb (99.999%) was deposited on the Ag (111) surface from a homemade Knudson cell. The Ag (111) substrate was held at 375 K during deposition and immediately annealed to 550 K for 60 min in the preparation chamber. The amount of Sb deposited on the surface was calibrated by counting the numbers of Sb atoms on a clean Ag (111) surface held at 200 K (see Appendix A for details). The sample was transferred *in vacuo* into the STM chamber for low-temperature STM at 77 K. A chemically etched tungsten tip was cleaned by $Ar^+$ sputtering prior to all measurements.

The first-principles calculations were performed under the framework of density functional theory (DFT) by using a Vienna ab initio simulation package (VASP) [15]. The projector augmented plane wave (PAW) method was applied to describe the ion-electron interaction [16]. The exchange-correlation terms were described by the Perdew-Burke-Ernzerhof (PBE) [17] functional. The energy cutoff of 300 eV was employed for the plane-wave basis sets. To model the surface system, a p(3×3×1) Ag (111) slab was used, which consisted of five atomic layers and a vacuum layer. The supercell was 8.67 x 8.67 x 26 Å in three dimensions (see Appendix B for details). The Brillouin zone was sampled with a gamma-centered mesh with a 3 x 3 x 1 k-point grid. During the geometry optimization, two Ag atomic layers at the bottom were fixed, whereas other Ag and Sb atoms were relaxed. The convergence criteria of the maximum force on each atom was 0.05 eV / Å, and the convergence criteria of the total energy was $10^{-5}$ eV.

## 3 Results and discussion

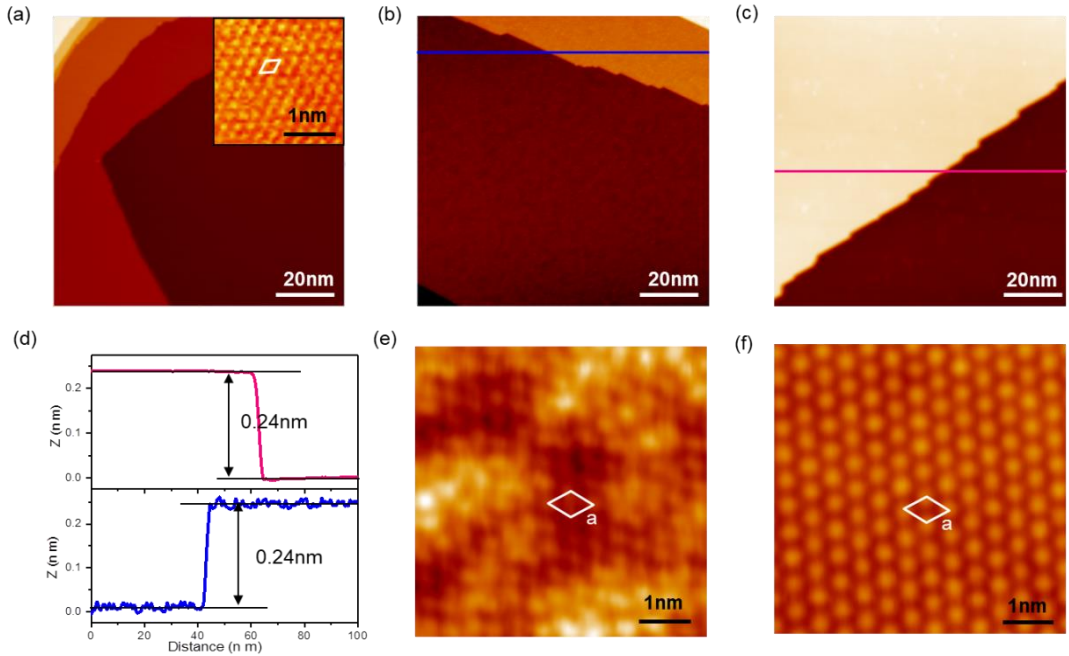

**Fig. 1.** Characterization of Ag (111) and Sb deposited on Ag (111). (a) STM topographic image of a clean Ag (111). (100 x 100 nm, $V_{bias}$ = −1.0 V, $I_t$ = 200 pA). Insert: atomically resolved STM image of the bare Ag (111) surface. (2.4 x 2.4 nm, $V_{bias}$ = 0.2 V, $I_t$ = 2 nA). (b) (100 x 100 nm, $V_{bias}$ = 0.5 V, $I_t$ = 500 pA) and (c) (100 x 100 nm, $V_{bias}$ = 0.5 V, $I_t$ = 500 pA) STM images of 0.9- and 1.8-ML Sb films deposited on Ag (111). (d) Line profiles along the blue and pink lines in (b) and (c). (e) (5.7 x 5.7 nm, $V_{bias}$ = −1.0 V, $I_t$ = 500 pA) and (f) (5.7 x 5.7 nm, $V_{bias}$ = −0.1 V, $I_t$ = 20 nA) high-resolution STM image of 0.9- and 1.8-ML films.

Fig. 1 shows STM results of Sb atoms deposited on Ag (111). The atoms of the clean Ag (111) surface are in a hexagonal arrangement with an in-plane lattice constant of 0.288 nm, as shown in the insert in Fig. 1a. Fig. 1b and 1c show large-area STM topographic images of two samples with 0.9- and 1.8-ML Sb (where 1 ML refers to 0.92 x $10^{15}$ atoms $cm^{-2}$, which is two-thirds of the atomic density of a Ag (111) surface) deposited on the Ag (111) surface, respectively. Both samples show flat terraces with a step height of 0.24 nm, and no island can be observed. High-resolution STM images (Fig. 1e and 1f) show that the topmost atoms of both samples are in a similar hexagonal arrangement and the in-plane lattice constants are both 0.50 nm. The unit cells of the topmost atoms are both 30° rotated with respect to the direction of underneath Ag atoms, suggesting a ($\sqrt{3} \times \sqrt{3}$)R30° reconstruction for both samples. For the sample with coverage of 0.9 ML, the surface shows high inhomogeneity, whereas a highly homogeneous surface is observed for the sample with coverage of 1.8 ML. These observations strongly suggest that the lower-coverage sample might be an alloy sample, whereas the higher-coverage sample is unlikely to be an alloy. To verify the possible constitution of the two samples, we monitored the initial alloy formation of submonolayer

Sb on Ag (111) deposited at a low temperature of approximately 200 K and annealed the sample at higher temperatures. The results are shown in Appendix A. It was shown that 0.2-ML Sb atoms deposited on Ag (111) held at 200 K form Sb chains, and no evidence of surface alloy was observed. The Sb chains fuse into small islands with a uniform $(\sqrt{3} \times \sqrt{3})R30°$ structure after the sample is annealed at room temperature for 11 h. Further annealing of the sample to 400 K caused noticeable inhomogeneity in the Sb islands. Because only two kinds of atoms are present on the surface, that is, Sb and Ag, this kind of inhomogeneity can only be ascribed to surface alloy with local uneven distributions. These observations strongly suggest that the surface of the sample in Fig. 1b and 1e should also be a surface alloy of Sb and Ag. Previous studies of Sb deposited on Ag (111) suggested a formation of $Ag_2Sb$ surface alloy, which also possesses a $(\sqrt{3} \times \sqrt{3})R30°$ structure [18, 19]. However, this alloy is described by ordered substitution of Sb for one-third of the surface Ag atoms, and the corrugation of this surface is fairly small (approximately 0.1 Å) [20]. This is not consistent with the height of the islands (2.2 Å) observed in Fig. A1e and A1f. A different model of surface alloy should be considered for the observed surface structures in this study. Moreover, the high homogeneity of the surface of the sample in Fig. 1c and 1f excludes the possibility of a surface alloy, meaning that the monoelemental Sb layer sitting atop the surface alloy is the most possible structure for the sample. If there are intercalating Ag atoms in the topmost layer, it should be observed as local inhomogeneity at certain sample biases with STM. Therefore, the possibility of the Ag intercalation into the topmost antimonene layer can be ruled out, as we did not find significant inhomogeneity in the antimonene at any sample bias.

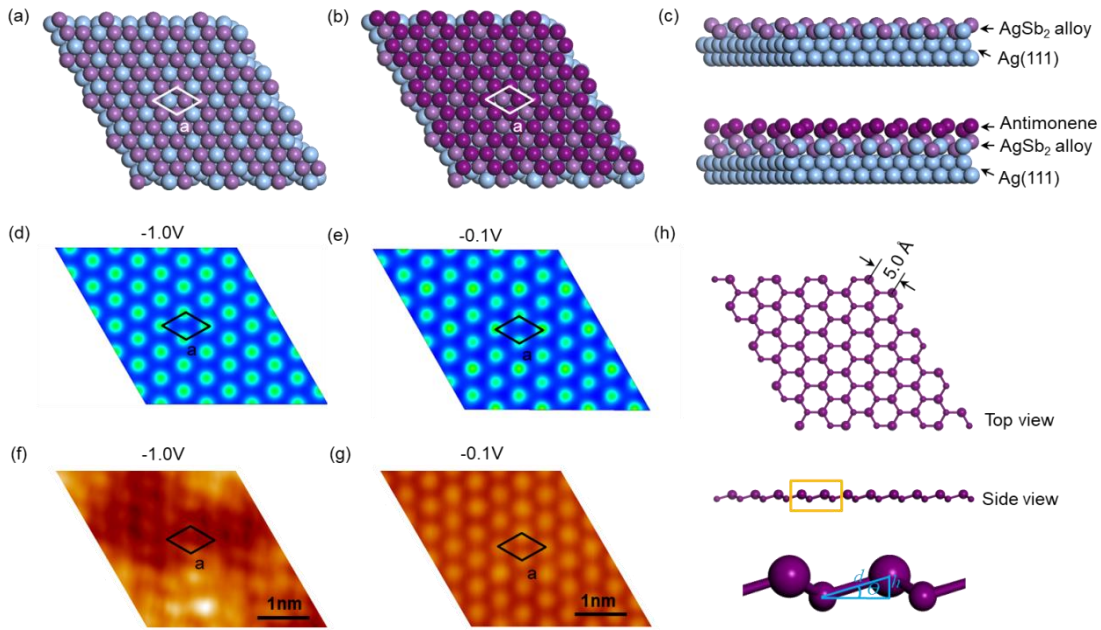

**Fig. 2.** Structural models (side and top views) and their simulated STM images compared with experiments. (a) Structural model, (d) simulated STM image, and (f) experimental STM image ($V_{bias}$ = -1.0 V, $I_t$ = 500 pA) of $AgSb_2$ surface alloy. (b) Structural model, (e) simulated STM image, and (g) experimental STM image ($V_{bias}$ = -0.1 V, $I_t$ = 20 nA) of a single layer of antimonene on a $AgSb_2$ surface alloy. (c) Side views of (a) and (b). (h) Top and side views of the atomic structure of the topmost antimonene layer. The bottom panel shows the buckling parameters.

To better understand our experimental results, we performed DFT calculations for various levels of thickness of Sb deposited on a Ag (111) surface. Details of the calculations are provided in Appendix B. Fig. 2 shows two optimal structural models with coverages of 1 and 2 ML Sb, respectively, as well as the simulated STM images compared with experimental results. The optimized structural model for 1-ML coverage shows a layer of AgSb$_2$ surface alloy epitaxially grown on the Ag (111) (Fig. 2a). Interestingly, this kind of arrangement can be regarded as a single layer of buckled antimonene intercalated with single Ag atoms into every honeycomb hollow. This arrangement leads to a high surface atomic corrugation (see Fig. 2c for a side view) that should be observed in STM images as a ($\sqrt{3} \times \sqrt{3}$)R30° hexagonal lattice. We found that the simulated STM image (Fig. 2d) of this surface alloy effectively reproduces the experimental image (Fig. 2f) despite certain local inhomogeneity. This kind of inhomogeneity can be understood by local uneven arrangement of Sb and Ag atoms. Considering the high atomic corrugation together with the coverage of the Sb atoms (0.9 ML) deposited on the Ag (111) surface, we conclude that the structural model of a AgSb$_2$ surface alloy rather than the Ag$_2$Sb surface alloy is the most possible one for the observed surface structure of the sample in Fig. 1b. The optimal surface structure for the higher-coverage sample in Fig. 1c is also calculated as shown in Fig. 2b. A monoelemental Sb layer of a buckled honeycomb structure (i.e., antimonene) is found to be stably grown atop the Ag$_2$Sb surface alloy. The simulated STM image (Fig. 2d) of the antimonene epitaxially grown on the AgSb$_2$ surface alloy is presented in Fig. 2e, which agrees well with the STM image of the corresponding sample with Sb coverage of 1.8 ML (Fig. 2g). More strikingly, we found that this kind of antimonene layer has an unprecedentedly large lattice constant of 5.0 Å, which is 21.4% larger than the predicted lattice constant of 4.12 Å for freestanding antimonene [8, 21, 22]. This kind of highly strained lattice is found to be stabilized with a lower-buckled structure of the epitaxially grown antimonene. The atomic structure and structural parameters for the strained buckled honeycomb structure is shown in Fig. 2h. The bond length ($d$) is determined to be 3.02 Å, which is 4.4% larger than that in freestanding antimonene (2.89 Å). The buckling height ($h$) and buckling angle ($\theta$) are 0.92 Å and 17.7° respectively, both of which are largely reduced compared to those for freestanding antimonene (1.30 Å and 26.8°). Interestingly, we noticed that a recent work on a Sn thin film grown on Ag (111) also reported the formation of stanene atop a Sn-Ag surface alloy [23].

Recent theory predicted that antimonene can be tuned to a 2D topological insulator by reducing the buckling height of the lattice under tensile strain [8]. The strain drives a band inversion in the vicinity of the Fermi level, which opens a bulk band gap due to spin-orbital coupling. The gap can be as large as 270 meV for tensile strain up to 18%. This bulk band gap allows antimonene to be a promising candidate material for realizing the QSH effect at high temperatures. In our work, the antimonene grown on the AgSb$_2$ alloy layer experiences high tensile strain that has not been experimentally realized in antimonene grown on previous substrates.

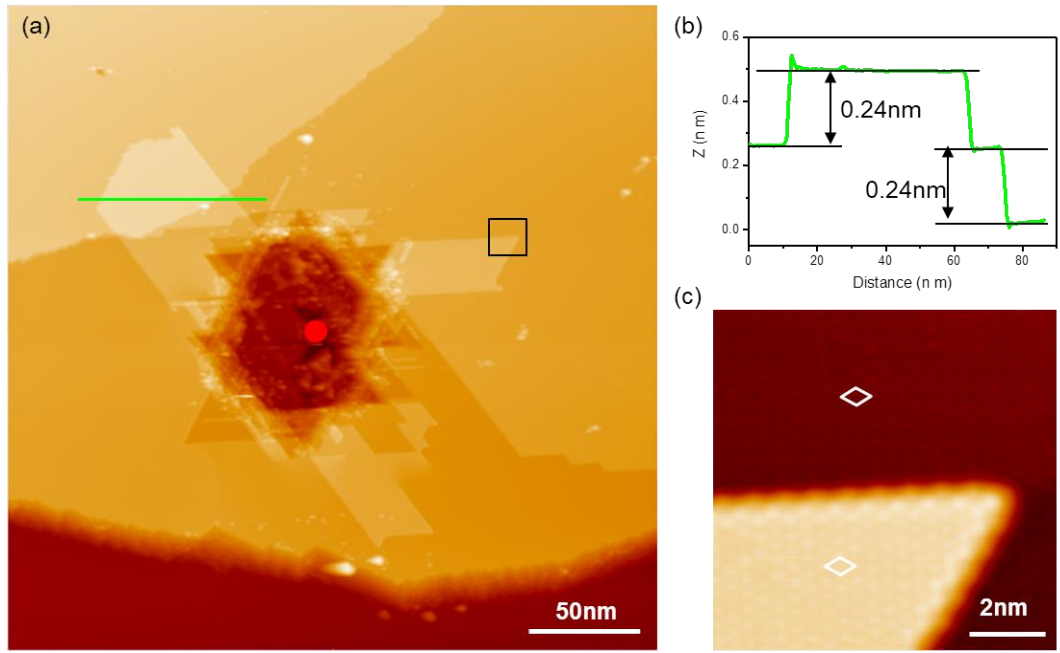

**Fig. 3.** Tip manipulation of the 1.8-ML sample when applying high-voltage pulse. (a) Large-scale STM image (292 x 292 nm, $V_{bias}$ = 0.5 V, $I_t$ = 1 nA) of a terrace after applying a 9-V pulse at the red point. (b) Line profile along the green line in (a). (c) Zoomed-in image of the black square in (a), (9 x 9 nm, $V_{bias}$ = −1.0 V, $I_t$ = 20 nA).

To further confirm that the surface of the sample in Fig. 1c is truly covered by a single layer of antimonene, we performed tip manipulation on the surface. Based on the calculations, the interaction between the antimonene layer and $AgSb_2$ alloy was approximately 0.5 eV/atom, if they were peeled off from their original positions. If the topmost layer is monolayer antimonene as shown in the theoretical model in Fig. 2b, it should be weakly coupled to the surface alloy underneath. Therefore, partly lifting the antimonene layer with the tip should be possible by applying high voltage pulses. In Fig. 3, we show that the top layer of the sample could actually be peeled off by applying a high voltage pulse with the tip. Fig. 3a shows an STM image of the sample after applying a 9-V pulse on the spot indicated by a red dot. Several pieces of the topmost layer were generated by the voltage pulse and flipped in the vicinity of the pit produced by the voltage pulse. The height of these pieces was 0.24 nm, equaling the step height in Fig. 1c. The atomic structure as well as the lattice constant of the flipped layers were the same in the underneath surface (Fig. 3c). Moreover, we can see that one of the flipped pieces lay across several steps, indicating its layered nature. The possibility of an alloy layer atop the sample is thus excluded. All these observations confirm the proposed structure in Fig. 2b and provide solid evidence that an antimonene layer exists on this sample.

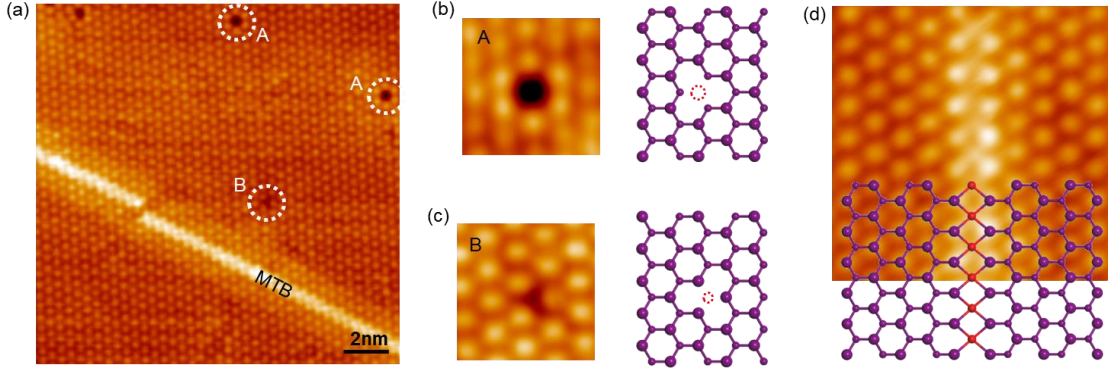

**Fig. 4.** STM topographic images and schematics of various kinds of defects. (a) Large-scale STM image (16 x 16 nm, $V_{bias}$ = 0.5 V, $I_t$ = 2 nA) showing several defects and a domain boundary. (b) (2.1 x 2.1 nm, $V_{bias}$ = −0.5 V, $I_t$ = 2 nA) and (c) (2.1 x 2.1 nm, $V_{bias}$ = 0.1 V, $I_t$ = 20 nA) are an atomically resolved STM image and corresponding schematic structures of two typical defects, Sb-A and Sb-B, marked in (a) with white dashed circles. (d) Atomically resolved STM image (4.4 x 4.4 nm, $V_{bias}$ = −0.1 V, $I_t$ = 20 nA) of the MTB with a schematic structure. The atoms between the two domains are denoted by red balls.

The buckled structure of the antimonene layer has two Sb atoms in a unit cell: one is in the upper plane (Sb-A); the other is in the lower plane (Sb-B). Therefore, there should be two primary point defects, which we did in fact observe. Fig. 4a is an STM image of the 1.8-ML sample showing several point defects and a line defect. Two main kinds of point defects exist, as shown in the high-resolution images of Fig. 4b and 4c, which can be attributed to the vacancies of missing Sb-A and Sb-B atoms, respectively. A schematic structure of these two defects is shown and compared with the STM observations. We also found a line defect in the sample, which can be identified as a mirror twin boundary (MTB). The high-resolution STM image of the MTB with its corresponding schematic structure partly superimposed on it is shown in Fig. 4d. The MTB is a one-dimensional topological defect consisting of a pair of distorted hexagons and an in-between Sb atom periodically repeated along the dislocation line. More interestingly, the in-between Sb atoms (red balls in Fig. 4d) were found to bond with four neighboring Sb atoms. This kind of configuration of fourfold coordination has not been observed previously in group-IV Xenes.

## 4  Conclusion

By using scanning tunneling microscopy combined with DFT calculations, we demonstrated in this study the successful synthesis of a highly strained antimonene on a surface alloy of $AgSb_2$ on a Ag (111) surface. The growth of antimonene with high tensile strain was facilitated by a $AgSb_2$ surface alloy as a template layer, which has a $(\sqrt{3} \times \sqrt{3})R30°$ hexagonal lattice with respect to the underneath Ag (111) surface. Our work showed for the first time that tensile strain higher than 20% can indeed be realized in antimonene epitaxially grown on a substrate with a large lattice mismatch, which may stimulate further investigation of possible QSH in similar epitaxial systems.

**Acknowledgements**

This work was supported by the National Key R&D Program of China (2016YFA0200603, 2017YFA0205004), the "Strategic Priority Research Program" of CAS (XDB01020100), the National Natural Science Foundation of China (Grants nos. 91321309, 21421063, 21473174), and the Fundamental Research Funds for the Central Universities (No. WK2060190027, WK 2060190084). A.-D.Z. acknowledges a fellowship from the Youth Innovation Promotion Association of CAS (2011322).

## Appendix A  Annealed temperature dependence of Sb film

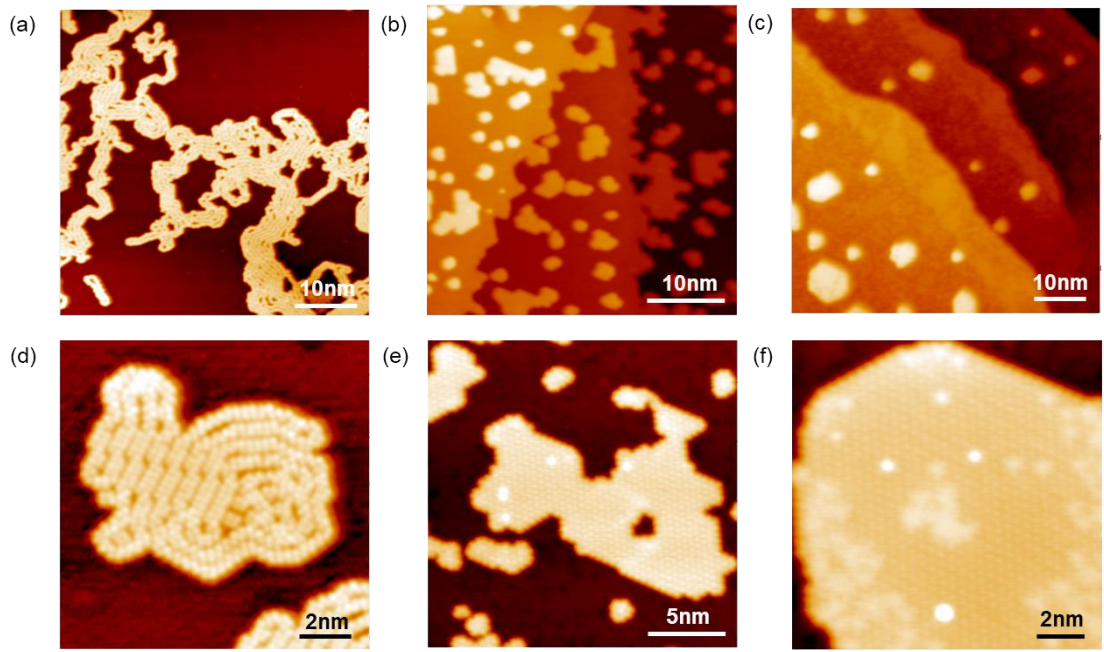

**Fig. A1.** Sequential STM topographical images showing the annealing-temperature dependence of 0.2-ML Sb deposited on Ag (111) at approximately 200 K. (a) (50 x 50 nm, $V_{bias}$ = −1.0 V, $I_t$ = 500 pA) and (d) (12 x 12 nm, $V_{bias}$ = −0.3 V, $I_t$ = 500 pA) are STM images of Sb atoms deposited at 200 K. The Sb atoms that have aggregated into nanochains and undecomposed $Sb_4$ molecules are still visible. No signature of alloy is evident. (b) (40 x 40 nm, $V_{bias}$ = −1.0 V, $I_t$ = 500 pA) and (e) (20 x 20 nm, $V_{bias}$ = −0.05 V, $I_t$ = 20 nA) STM image of the sample annealed at 300 K for 11 h. Sb nanochains and $Sb_4$ molecules disappear and small islands appear. The height of the island is 2.2 Å, which is lower than the Ag step height of 2.4 Å. This value cannot be explained by the $Ag_2Sb$ alloy model in which both Sb and Ag atoms are in a planar hexagonal lattice, but can be understood with the $AgSb_2$ alloy model in which the alloy layer is highly buckled. (c) (60 x 60 nm, $V_{bias}$ = 3.0 V, $I_t$ = 600 pA) and (f) (13 x 13 nm, $V_{bias}$ = −0.02 V, $I_t$ = 5 nA) STM images of the sample subjected to further annealing at 400 K for 1 h. The islands have the same structure and height as those in (e) but with some inhomogeneity, which can be understood by the uneven arrangement of Sb and Ag atoms in the alloy.

## Appendix B  Details of theoretical calculations and simulations

First, single Sb atom adsorption on Ag (111) was considered for the low-coverage limit. After geometry optimization, the Sb atom could be stable on either hollow (Fig. A2a) or bridge (Fig. A2b) sites. However, it was not stable on the top site, where Sb could replace Ag on the surface spontaneously during the geometry optimization (Fig. A2c). Then, the buckled antimonene was placed on Ag (111) on hollow and top sites (Fig. A2e) to simulate the monolayer case. After relaxation, the antimonene either turned out to be graphene-like with a honeycomb lattice on hollow sites (Fig. A2d) or exchanged with a Ag atom in the substrate to become disordered (top site case, Fig. A2e). In addition, those two structures did not match the experimental observations. The $Ag_2Sb$ model was thus tested as the initial substrate, but the Sb atoms in the surface alloy would be extracted when a submonolayer Sb atom adsorbed on the surface (Fig. A2f). In addition, this model still did not match the initial growth process as observed in Appendix A. Finally, a $AgSb_2$ model (Fig. 2a) was found to match the experimental observation very well and was used for further calculations. The antimonene and $AgSb_2$ layers showed large difference in terms of electronic structure. The calculated band structures for both are shown in Fig. A3.

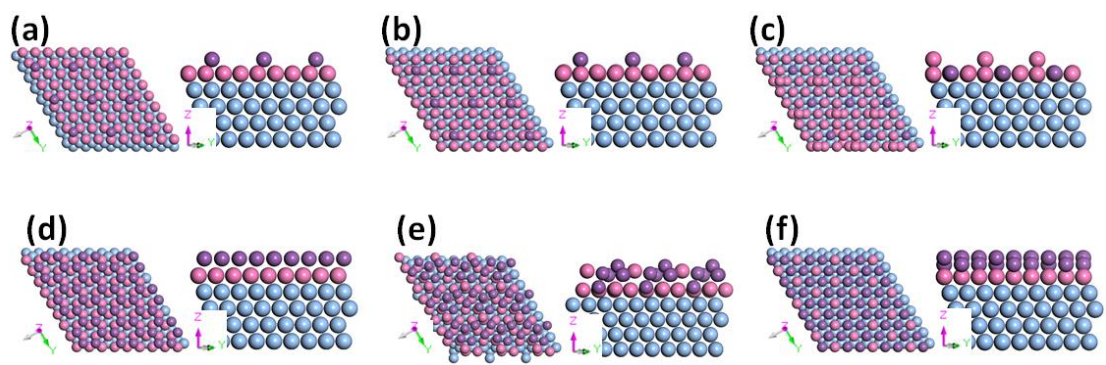

**Fig. A2.** Adsorption of a single Sb atom on Ag (111) including (a) hollow, (b) bridge, and (c) top sites. The structure of antimonene placed on Ag (111) on (d) hollow and (e) top sites. (f) $Ag_2Sb$ alloy model with Sb adsorption on the surface. Blue and pink balls represent Ag atoms, and purple balls represent Sb atoms.

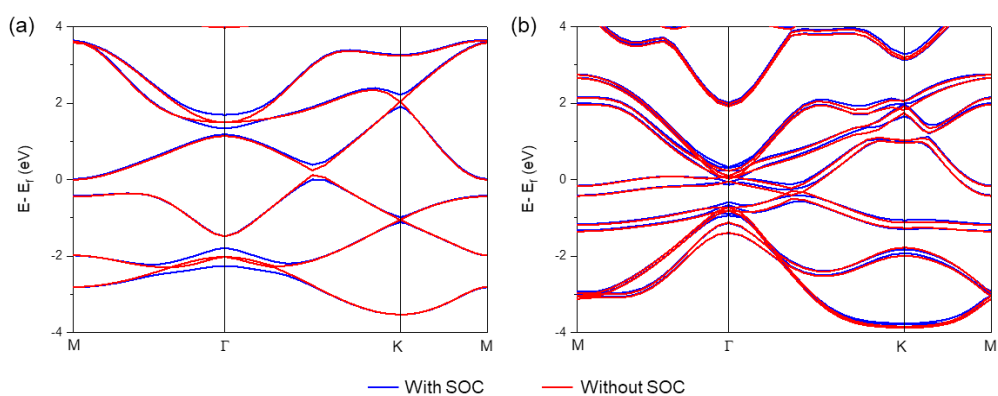

**Fig. A3.** Calculated band structures for the antimonene (a) and $AgSb_2$ (b) layers. The substrate atoms are removed in the calculation for better visualization of their intrinsic electronic structures.